\documentclass[twocolumn,epjc3]{svjour3}
\usepackage{amsfonts}
\usepackage{amsmath}
\usepackage{amssymb,bbold}
\usepackage[normalem]{ulem}
\usepackage{booktabs}
\usepackage{cancel}
\usepackage{bbm}
\RequirePackage[T1]{fontenc}
\smartqed
\RequirePackage{graphicx}
\RequirePackage{mathptmx}
\RequirePackage{flushend}
\RequirePackage[numbers,sort&compress]{natbib}
\RequirePackage[colorlinks,citecolor=blue,urlcolor=blue,linkcolor=blue]{hyperref}
\RequirePackage{color}

\begin{document}

\title{Ground state of a bosonic massive charged particle in the presence of external fields in a G\"{o}del-type spacetime}
\author{Edilberto O. Silva\thanksref{e1}}
\thankstext{e1}{e-mail: edilbertoo@gmail.com}
\institute{Departamento de F\'{i}sica, Universidade Federal do
           Maranh\~{a}o, 65080-805, S\~{a}o Lu\'{i}s, Maranh\~{a}o,
           Brazil
           } \journalname{Eur. Phys. J. C}
\date{Received: date / Accepted: date}
\maketitle

\begin{abstract}
The relativistic quantum dynamics of a spinless charged particle interacting with both Aharonov-Bohm and Coulomb-type potentials in the G\"{o}del-type spacetime is considered. The dynamics of the system is governed by the Klein-Gordon equation with interactions. We verify that it is possible to establish a quantum condition between the energy of the particle and the parameter that characterizes the vorticist of the spacetime. We rigorously analyze the ground state of the system and determine the corresponding wavefunctions to it.
\end{abstract}

\section{Introduction}
\label{sec:introduction}

The relativistic quantum dynamics of spinless particles in the presence of external fields has been an object of study for many decades. The physical properties of the systems are accessed by the solution of the Klein-Gordon field equation with electromagnetic interactions \cite{Book.2000.Greiner,Book.Landau.Vol4}. The electromagnetic interactions are introduced into the Klein-Gordon equation through the so called minimal substitution, $p^{\mu}\rightarrow p^{\mu}-eA^{\mu}$, where $e$ is the charge and $A^{\mu}$ is the four-potential of the electromagnetic field. It is also known that interactions can be implemented by making a modification in the mass term as ~$M\rightarrow M+S\left(\mathbf{r},t\right)$, where $S\left(\mathbf{r}, t\right)$ is a scalar potential. Contrary to the minimal substitution, the scalar interaction $S\left(\mathbf{r},t\right)$ is independent of the charge of the spinless particle considered and, consequently, it has the same effect on particles and antiparticles, respectively. In the context of recent applications, such couplings have been used to study the motion of particles in external fields in various branches of physics. For example, in Ref. \cite{PRD.2016.93.045033}, the minimal substitution was used to analyze the particle scattering and vacuum instability problems in a constant electric field for the special set of stationary solutions of the Dirac and Klein-Gordon equations. A procedure to construct generalized ladder operators for the Klein-Gordon equation with a scalar curvature term was obtained in Ref. \cite{PRD.2018.97.025011}. Another interesting work addresses the study of nonlinear quantum electrostatic waves in a pseudorelativistic quantum semiconductor plasm \cite{PRE.2015.91.043108}. In this description, the authors considered the substitution $p^{0}\rightarrow p^{0}-eA^{0}$ and $p\rightarrow p$, and they showed that the system is governed by the Klein-Gordon equation for the collective wave functions of the conducting electrons and Poisson's equation for the electrostatic potential. It is also common to use both the scalar and the minimum couplings to study physical models in gravitation. In this direction, we call attention to the relativistic quantum mechanics of particles in external fields in the cosmic string spacetime \cite{Book.2000.Vilenkin,EPJC.2012.72.2051,MPLA.2018.33.1850025,PRD.2014.90.125014,PRD.2010.82.084025,EPJC.2018.78.13,EPJC.2016.76.61}. In general relativity, these couplings appear in the study of the quantum dynamics of relativistic particle in the G\"{o}del-type
spacetime \cite{EPJPlus.2015.130.36}. The G\"{o}del spacetime is considered a usual framework for studying physical systems in general relativity \cite{RMP.1949.21.447}. From the topological point of view, this spacetime is known to be geodesically complete and it is a singularity-free solution of the Einstein field equation \cite{JHEP.2012.2012.32,PRD.2013.87.087503,PRD.2015.92.123541,EPJC.2017.77.289}.
The interest in this issue led some researchers to propose a generalization of this universe by examining the conditions for spacetime homogeneity of a Riemannian manifold with a G\"{o}del-type metric \cite{PRD.1983.28.1251}. The study of physical systems involving G\"{o}del-type solutions is a problem of current interest. The generalization of Ref. \cite{PRD.1983.28.1251} resulted in the first exact G\"{o}del-type solution of the Einstein equation for a rotating universe, including the problem of causality in (Raychaudhuri-Thakurta)-homogeneous Riemannian manifolds \cite{PRSLA.1968.304.81}. In this context, several important contributions have been made, as can be see in Refs. \cite{JMP.1999.40.4011,GRG.2008.40.2115,JCAP.2004.2004.012,EPJC.2014.74.2935}.
In this paper, we study the relativistic quantum dynamics of a spinless
charged particle interacting with both the Aharonov-Bohm and the Coulomb-type potentials in the G\"{o}del-type spacetime, with line element in cylindrical coordinates written as (in natural units, $\hbar=c=G=1$)
\begin{equation}
ds^{2} = -\left(dt+\Omega \frac{\sinh ^{2}lr}{l^{2}}d\phi \right)^{2}+\frac{\sinh^{2}2lr}{4l^{2}}d\phi^{2}+dr^{2}+dz^{2},  \label{Gmetric}
\end{equation}
with $0\leq r<\infty ,0\leq \phi \leq 2\pi$ and $-\infty <(z,t)<\infty$. The metric (\ref{Gmetric}) is characterized by the parameters $(l,\Omega )$, both with the dimensions of the inverse length, with $\Omega$
representing the vorticity of the spacetime. The physics contained in the
metric (\ref{Gmetric}) is related to the problem of a charged particle in
a plane in the presence of a constant magnetic field, with $\Omega$ playing the role of such a magnetic field \cite{JCAP.2004.2004.012}. We address the particular case where the limit $l\rightarrow 0$ is imposed for the metric (\ref{Gmetric}), which reveals that the resulting metric has the same geometry as that of the so called Som-Raychaudhuri spacetime \cite{PRSLA.1968.304.81}. In our approach, the relevant equation is the
Klein-Gordon equation with interactions. We solve this equation and
determine the energy eigenvalues and the corresponding wave functions. We evaluate the
expression for the energy spectrum of the particle and establish conditions for its validity. Finally, we consider the fundamental state of the system and study it in detail.

\section{The Klein-Gordon equation}
\label{KG}

After imposing the limit $l\rightarrow 0$ in (\ref{Gmetric}), the metric is simplified and it takes the form
\begin{equation}
ds^{2}=-\left( dt+\Omega r^{2}d\varphi \right) ^{2}+r^{2}d\varphi
^{2}+dr^{2}+dz^{2}.  \label{metric}
\end{equation}
The relativistic quantum dynamics of a spinless charged particle of mass $M$ in an electromagnetic field is described by the Klein-Gordon equation with minimal coupling, which is written in the covariant form as
\begin{equation}
\left[\frac{1}{\sqrt{-g}}D_{\mu}\left(\sqrt{-g}g^{\mu \nu}D_{\nu}\right)-M^{2}\right] \psi \left(t, r, \varphi, z\right) =0, \label{kg}
\end{equation}
with $D_{\mu}=\partial_{\mu}+ieA_{\mu}$, where $e$ is the electric
charge, $A_{\mu}$ denotes the quadrivector potential associated to the
electromagnetic field and $g=\det \left(g^{\mu \nu}\right) $ is the
determinant of the metric tensor in the spacetime of the line element (\ref{metric}). We can exploit the translational symmetry by considering the Klein-Gordon equation for the case where $\partial /\partial z=0$ and $A_{z}=0$. The Aharonov-Bohm potential is given by $eA_{\varphi}=-\Phi /r$, where $\Phi =\Phi_{B}/\left(2\pi /e\right)$ is the magnetic flux parameter. In order to include the Coulomb potential, we put $eA_{0}=V\left(r\right) =\xi /r$, where $\xi = \pm e^2$ (the choice of sign corresponds to either repulsion or attraction, respectively). Equation (\ref{kg}) assumes the form
\begin{align}
\Bigg\{\frac{\partial^{2}}{\partial r^{2}}+\frac{1}{r}\frac{\partial}{\partial r}& +\frac{1}{r^{2}}\left[\frac{\partial}{\partial \varphi}
+i\Phi -\Omega r^{2}\left(\frac{\partial }{\partial t}+\frac{i\xi}{r}
\right) \right]^{2}  \notag \\
&-\left(\frac{\partial}{\partial t}+\frac{i\xi }{r}\right)^{2}-M^{2}
\Bigg\}\psi \left(t,r,\varphi \right) =0,  \label{KGa}
\end{align}
It is known that the partial wave functions for the Aharonov-Bohm problem can be obtained exactly even though a Coulomb potential is included into the equation of motion \cite{JPA.1992.25.L183}. Then, for the wave function $\psi \left(t,r,\varphi \right) $, we make the following ansatz:
\begin{equation}
\psi \left(r,\varphi, t\right) =e^{-iEt+im\varphi}R\left(r\right),\label{ansatz}
\end{equation}
with $m=0,\pm 1,\pm 2,\ldots$. This leads to the radial equation
\begin{equation}
\left[\frac{d^{2}}{dr^{2}}+\frac{1}{r}\frac{d}{dr}-\frac{j^{2}}{r^{2}}-
\frac{a}{r}+br-\omega^{2}r^{2}+k^{2}\right] R\left(r\right) =0,
\label{eqh}
\end{equation}
where $j^{2}=\left(m+\Phi \right)^{2}-\xi ^{2}$, $a=2\xi \left[E-\Omega \left(m+\Phi \right) \right]$, $b=2E\xi \Omega^{2}$, $\omega^{2}=E^{2}\Omega^{2}$ and $k^{2}=E^{2}-M^{2}-\xi^{2}\Omega^{2}-2\Phi \Omega E-2m\Omega E$. In order to solve the differential equation (\ref{eqh}), it is convenient to define a new radial function,
\begin{equation}
R\left(r\right) =\frac{f\left(r\right)}{\sqrt{r}},
\end{equation}
which, together with the variable change $x=\sqrt{\omega}r$, leads to the following Schr\"{o}dinger-type equation:
\begin{equation}
\left[\frac{d^{2}}{dx^{2}}-\frac{\left(j^{2}-\frac{1}{4}\right)}{x^{2}}-
\frac{a^{\prime}}{x}+b^{\prime}x-x^{2}+\frac{k^{2}}{\omega}\right]
f\left(x\right) =0, \label{Schr}
\end{equation}
where $a^{\prime}={a/}\sqrt{\omega}$ and $b^{\prime}=b/\sqrt{\omega^{3}}$. Now, let us study the asymptotic limits of Eq. (\ref{Schr}). This
analysis reveals that it is possible to obtain normalizable eigenfunctions by writing $f\left(x\right)$ as
\begin{equation}
f\left(x\right) ={x}^{j+\frac{1}{2}}e{^{-\frac{1}{2}\,x\left(x+b^{\prime}\right)}y}\left(x\right),  \label{solution}
\end{equation}
where ${y}\left( x\right)$ is a new function to be determined. Substituting
Eq. (\ref{solution}) into the Eq. (\ref{Schr}), we get
\begin{align}
x\frac{d^{2}y}{d{x}^{2}}& +\left(1+\alpha -\beta x-2x^{2}\right) \frac{dy}{d{x}}  \notag \\
& +\left\{\left(\gamma - \alpha -2\right) x-\frac{1}{2}\left[\delta
+\left(1+\alpha \right) \beta \right] \right\} y=0,  \label{Heun-equation}
\end{align}
where $\alpha =\,{2{j}}$, $\beta =-b^{\prime }$, $\gamma =\frac{k^{2}}{\omega}+\frac{1}{4}b^{\prime 2}$ and $\delta =2a^{\prime}$. Equation (\ref{Heun-equation}) is the biconfluent Heun equation in its canonical form \cite{ronveaux1995heun,JCAM.37.161.1991}. This equation has a regular singularity at $x=0$, and an irregular singularity at $\infty $ of rank $2$. The use of the Frobenius method allows us to determine a series solution for (\ref{Heun-equation}). Upon writing
\begin{equation}
N\left(\alpha ,\beta ,\gamma ,\delta ;x\right) =\sum\limits_{\upsilon
=0}^{\infty}\frac{\mathcal{A}_{\upsilon }\left(\alpha ,\beta ,\gamma
,\delta \right)}{\left(1+\alpha \right)_{\upsilon}}\frac{x^{\upsilon}}{\upsilon !}, \label{Hs}
\end{equation}
where
\begin{equation}
\left(1+\alpha \right)_{\upsilon}=\frac{\Gamma \left(\upsilon+\alpha+1\right)}{\Gamma \left(\alpha+1\right)},
\end{equation}
we obtain the three-term recurrence relation,
\begin{align}
\mathcal{A}_{\upsilon+2}& =\left\{\left(\upsilon +1\right) \beta +\frac{1}{2}\left[\delta+\beta \left( 1+\alpha \right) \right] \right\} \mathcal{A}_{\upsilon+1}  \notag \\
& -\left(\upsilon +1\right) \left(\upsilon +1+\alpha \right) \left[\gamma-2-\alpha-2\upsilon \right] \mathcal{A}_{\upsilon},  \label{recur}
\end{align}
with
\begin{equation}
\mathcal{A}_{0}=1,
\end{equation}
\begin{equation}
\mathcal{A}_{1}=\frac{1}{2}\left[\delta +\beta \left(1+\alpha \right)
\right] .
\end{equation}
It can be verified directly in the recursion relation (\ref{recur}) that, if $\mathcal{A}_{\upsilon +2}=0$, the function $N\left(\alpha, \beta,\gamma, \delta ; x\right) $ becomes a polynomial of degree $n$. This can be accomplished if and only if the following conditions are imposed:
\begin{equation}
\gamma -\alpha -2=2n,  \label{cndA}
\end{equation}
and
\begin{equation}
\mathcal{A}_{n+1}=0,  \label{cndB}
\end{equation}
where $n$ is a positive integer. In this case, the $\left(n+1\right)$th
coefficient in the series expansion is a polynomial of degree $n$ in $\delta$. When $\delta$ is a root of this polynomial, the $\left(n+1\right)$th and the subsequent coefficients are identically null and the series truncates, resulting in a polynomial form of degree $n$ for series solution $N\left(\alpha ,\beta, \gamma, \delta; x\right)$. The general solution of (\ref{Heun-equation}) is given by
\begin{equation}
y\left(x\right)=c_{1}\,N\left(\alpha ,\beta, \gamma, \delta; x\right)
+c_{2}\mathrm{\,}x^{-\alpha }N\left(-\alpha, \beta, \gamma, \delta; x\right),  \label{solHeun}
\end{equation}
where $N\left( \alpha ,\beta ,\gamma ,\delta ;x\right) $ are the Heun
functions and $c_{1}$ and $c_{2}$ are normalization constants. The solution (\ref{solution}) becomes
\begin{align}
& f\left( x\right) =c_{1}\,{x}^{j+\frac{1}{2}}e{^{-\frac{1}{2}x\left(x+b^{\prime}\right)}}N\left({2{j}}, -b^{\prime}, \frac{k^{2}}{\omega}+\frac{1}{4}b^{\prime 2},2a^{\prime };x\right) \notag \\
&+c_{2}\,{x}^{-j+\frac{1}{2}}e{^{-\frac{1}{2}x\left(x+b^{\prime}\right) }}N\left(-{2{j}},-b^{\prime},\frac{k^{2}}{\omega}+\frac{1}{4}b^{\prime
2},2a^{\prime };x\right) .  \label{endsolution}
\end{align}
By using the conditions (\ref{cndA}) and (\ref{cndB}), we can determine an expression for the energy eigenvalues of the particle. From condition (\ref{cndA}), we obtain
\begin{equation}
E_{nm}^{2}-2\Omega \left( {n}+j+m+\Phi +{1}\right) E_{nm}-M^{2}=0{.}
\label{Eqenergy}
\end{equation}
By solving (\ref{Eqenergy}) for $E_{nm}$, we find
\begin{align}
E_{nm}& =\left( {n}+j+m+\Phi +{1}\right) \Omega  \notag \\
& \pm \sqrt{\Omega ^{2}\left({n}+j+m+\Phi +{1}\right) ^{2}+M^{2}}.
\label{eigen}
\end{align}
As we can see, the eigenvalues in Eq. (\ref{eigen}) are given in terms of all the physical parameters of the problem, that is, all the physical
quantities present in Klein-Gordon equation (\ref{KGa}). However, these
eigenvalues do not constitute the spectrum of the system. In order to make the Eq. (\ref{eigen}) to represent it, we need to implement the condition (\ref{cndB}). As pointed out in Ref. \cite{AoP.2014.347.130}, such condition (\ref{cndB}) gives rise to a quantum condition between the energy eigenvalues and the frequency of the model. In our case, the frequency is characterized by the vorticity $\Omega $ of the spacetime. On the other hand, in some particular models found in the literature, the authors suggest that this quantum condition has been used to "restore" some physical parameter of the problem which do not appear in the expression for the energy derived in Eq. (\ref{cndA}) (and which is present in the equation of motion) (see for example the Ref. \cite{EPJC.2012.72.2051}). In our case, the requirement (\ref{cndB}) appears as a necessary condition to ensure the existence of a set of energy eigenfunctions of the system. In fact, it is well-known of quantum mechanics that for a set of eigenvalues $E_{n}$ of a given hamiltonian operator $H_{n}$ , we have the set of energy eigenfunctions $\psi _{n}$ corresponding to each eigenvalue. The use of the condition (\ref{cndB}) to determine quantum conditions to restore physical parameters, a priori, is not a general rule. Each physical model should either require or not the establishment of such a condition.

Here, the requirement (\ref{cndB}) also provides a quantum condition between energy and frequency as in Ref. \cite{AoP.2014.347.130}. We would like to emphasize that, for $n\geqslant 1$, the equations resulting from conditions (\ref{cndA}) and (\ref{cndB}) are more complicated and they generally do not provide physically acceptable energy eigenvalues. For this reason, we investigate only the ground state of the particle. Thus, the $n=0$ state provides two quantum conditions, given by
\begin{align}
\left[\Omega \allowbreak_{0m}\right]_{1}& =\frac{2E_{0m}}{2\left(j+m+\Phi \right)+1},  \label{Omega1} \\
\left[\Omega \allowbreak_{0m}\right]_{2}& =\frac{2E_{0m}}{2\left(-j+m+\Phi \right) -1}.  \label{Omega2}
\end{align}
The final expressions for the energies are obtained after substituting (\ref{Omega1}) and (\ref{Omega2}) in (\ref{Eqenergy}):
\begin{eqnarray}
\left[ E_{0m}\right]_{1} &=&\pm \sqrt{-\frac{2\left( j+m+\Phi \right) +1}{2\left( j+m+\Phi \right) +3}}\,M,  \label{E0m1} \\
\left[ E_{0m}\right]_{2} &=&\pm {\sqrt{-\frac{2\left(-\,j+m+\Phi \right) -1}{2\left(3j+m+\,\Phi \right) +5}}\,M}.  \label{E0m2}
\end{eqnarray}
Thus, there exist solutions both for positive as well as for negative
energies, respectively, with the requirement that
\begin{align}
& \frac{2\left(j+m+\Phi \right) +1}{2j+2m+2\Phi+3}<0,  \label{cn1} \\
& \frac{2\left(-\,j+m+\Phi \right) -1}{2\left(3j+ m+ \Phi \right) +5}<0,\label{cn2}
\end{align}
to ensure that $\left[E_{0m}\right]_{1,2}$ are real. Evidently, for the ground state, the regular Heun function, Eq. (\ref{Hs}), becomes
\begin{equation}
N\left({2{j}}, -b^{\prime}, \frac{k^{2}}{\omega}+\frac{1}{4}b^{\prime 2},2a^{\prime };x\right)=1.
\end{equation}
In this case, the unnormalized bound state wave function for our problem corresponding to the regular solution at the origin is given by
\begin{equation}
f\left( x\right) ={x}^{j+\frac{1}{2}}e{^{-\frac{1}{2}x\left(x+b^{\prime}\right)}}.\label{functions}
\end{equation}
The conditions (\ref{cn1}) and (\ref{cn2}) reveal that the energies are physically acceptable only for appropriate values of $m$, $\Phi$ and $\xi $. It is important to note that, at least for the ground state, when $M=0$ in Eqs. (\ref{E0m1}) and (\ref{E0m2}), the energies are identically zero. This fact that does not occur for the spectrum of Ref. \cite{EPJC.2014.74.2935}, where $M=0$ leads immediately to the analogue Landau levels.
\begin{figure}[tbh]
\includegraphics[scale=0.8]{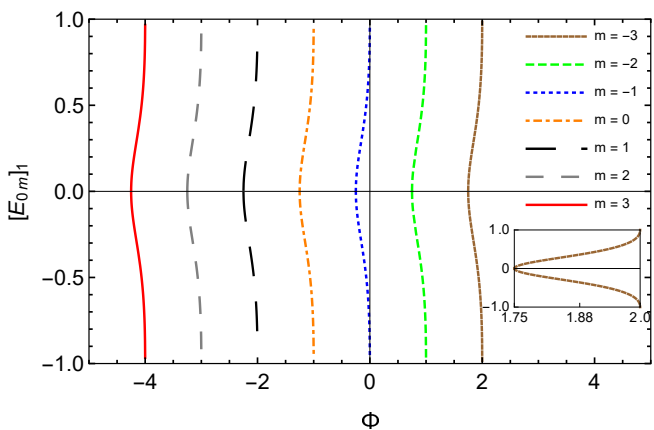}
\caption{Profile of the ground state (Eq. (\ref{E0m1})) as a
function of the magnetic flux $\Phi$ for $M=1$ and $\protect\xi =1$. The
energy is symmetric in relation to the quantum number $m$. The behavior for small variations of magnetic flux for $m=-3$ is shown by the inset.}
\label{Energy_freq1}
\end{figure}
\begin{figure}[tbh]
\begin{center}
\includegraphics[scale=0.8]{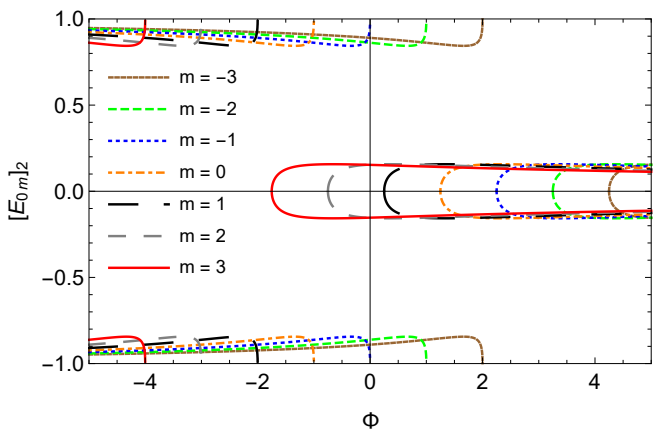}
\end{center}
\caption{Profile of the ground state (Eq. (\ref{E0m2})) as a
function of the magnetic flux $\Phi$ for $M=1$ and $\protect\xi=1$.
Clearly, we have regions in which energies are not defined for some values of $m$, existing only in small intervals.}
\label{Energy_freq2}
\end{figure}
The energy (\ref{E0m1}) is plotted as a function of the magnetic flux
parameter $\Phi $ in Fig. \ref{Energy_freq1} for some values from $m$. We
can see that both the particle and the antiparticle energy levels are members of the energy spectrum. Another explicit evidence in Fig. \ref{Energy_freq1} is that the energies are symmetrical about  $\left[E_{0m}\right]_{1}=0$ and, since the positive and the negative energies never intercept each other, we can see that there is no channel for spontaneous particle-antiparticle creation. Furthermore, we also see the presence of regions where energies are not allowed. The appearance of such regions is justified by the imposition of the quantum conditions (\ref{cn1}) and (\ref{cn2}). \begin{figure}[tbh]
\begin{center}
\includegraphics[scale=0.8]{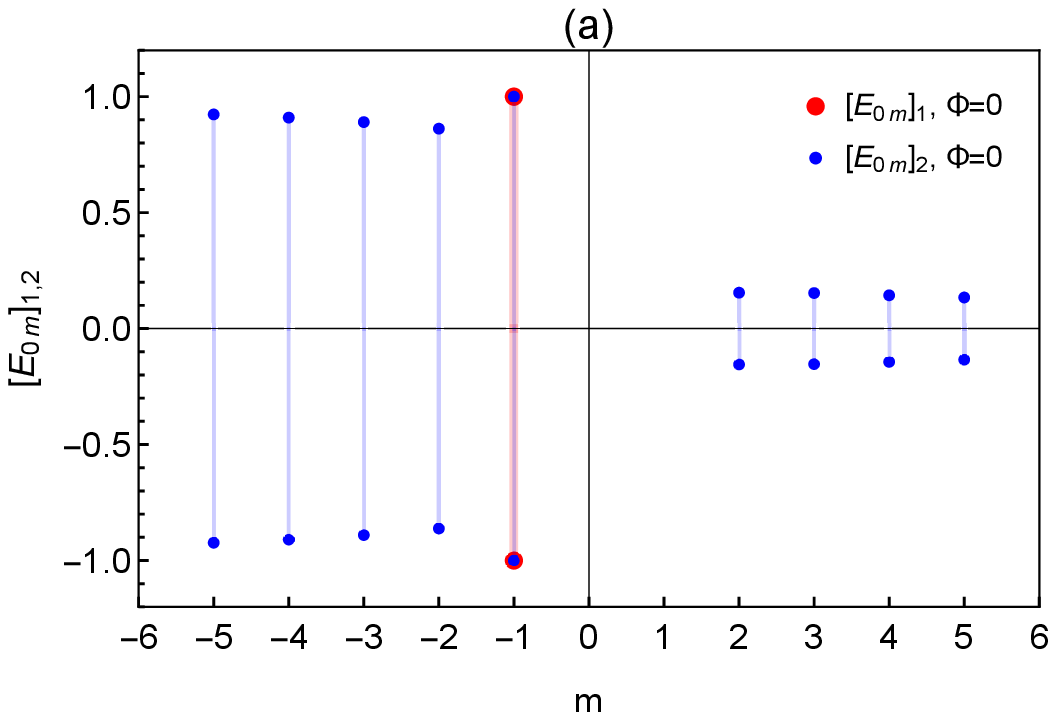}\\ \vspace{0.5cm}
\includegraphics[scale=0.8]{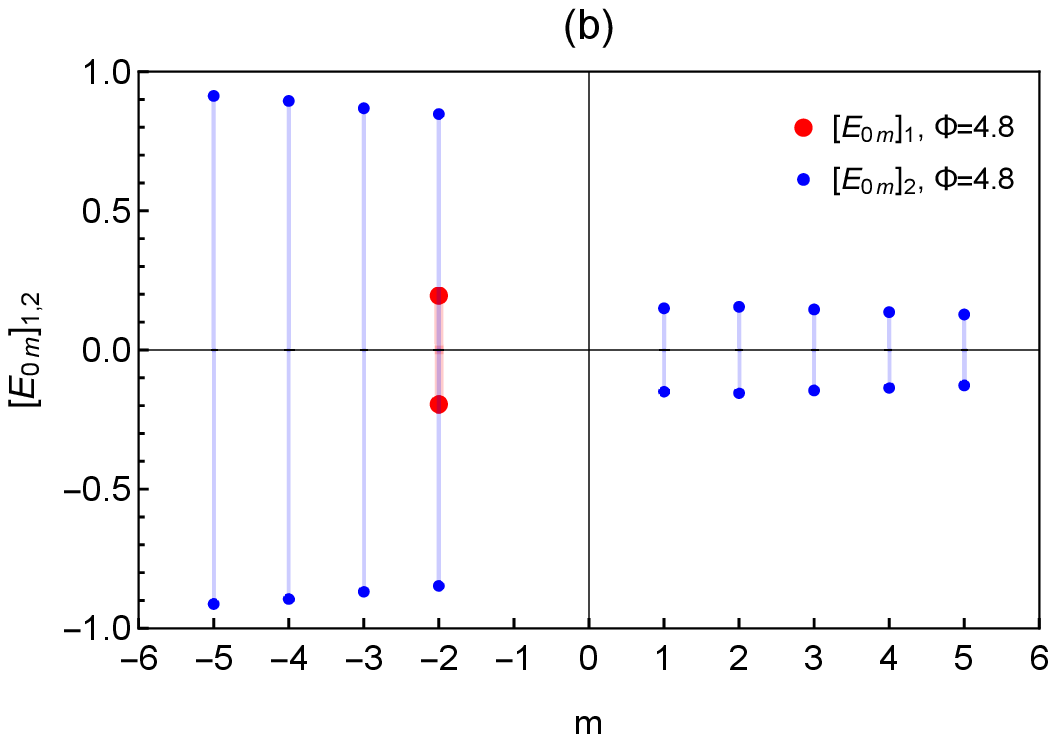}
\end{center}
\caption{Ground state $\left[E_{0m}\right]_{1}$ (blue dot) and $\left[E_{0m}\right]_{2}$ (dot red) as a function of the quantum number $m$ for $M=1$, $\protect\xi =1$ and $\Phi=0$ (Fig. (a)) and $\Phi=4.8$ (Fig. (b)).}
\label{Enery_null_const_flux}
\end{figure}
\begin{figure}[!h!]
\includegraphics[scale=0.6]{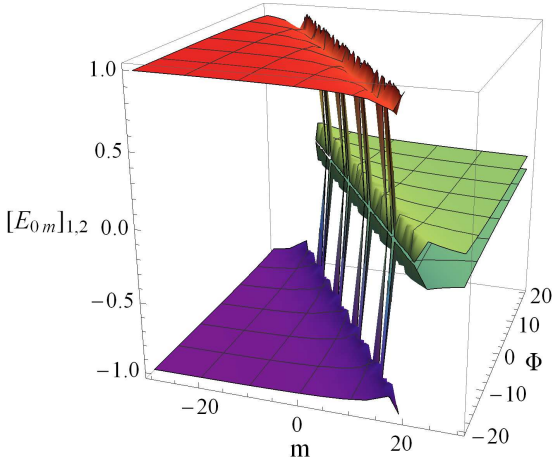}
\caption{Ground state $\left[E_{0m}\right]_{1,2}$ as a function of the
quantum number $m$ for $M=1$ and $\protect\xi =1$.}
\label{Energy3D}
\end{figure}
This feature also reveals that the energy is bounded in the finite interval $\left[E_{0m}\right]_{1,2}=\pm 1$. For positive values of $m$, including $m=0$, there is a magnetic flux inversion. We can also see that for small variations of magnetic field the energy is abruptly modified (see inset in Fig. \ref{Energy_freq1}). This occurs when the magnetic flux takes on integer values. This characteristic is manifested when the magnetic flux takes on integer values.
\begin{figure}[!h!]
\includegraphics[scale=0.8]{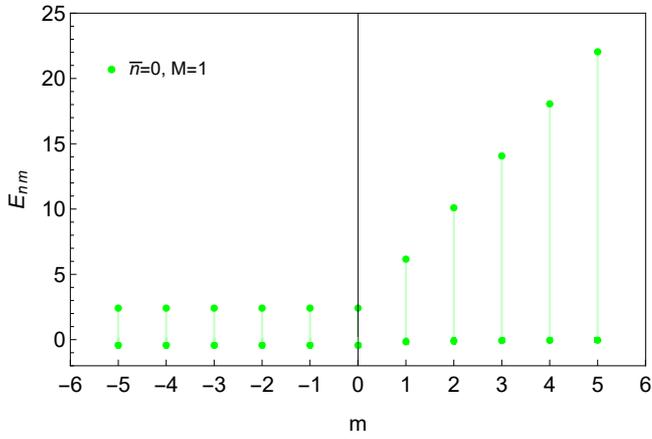}
\caption{Ground state (Eq. (\ref{Eq_Energy_Extern_Null})) as a function of the quantum number $m$ for $M=1$ and $\Omega =1$.}
\label{Energy_Extern_Null}
\end{figure}
The profile of the energy (\ref{E0m2}) is shown in Fig. \ref{Energy_freq2} for the same values of $m$ considered in Fig. \ref{Energy_freq1} and, it also exhibits the same characteristics. The only difference is that in Fig. \ref{Energy_freq2} appears new regions of bounded energies. We also investigate the ground state of the system by assuming a constant magnetic field  in comparison to the case when it is identically null (Fig. \ref{Enery_null_const_flux}). For this case, the behavior of it is similar to which is observed in Figs. \ref{Energy_freq1} and \ref{Energy_freq2}. To show the different effects of the Coulomb-type potential and Aharonov-Bohm flux on the ground state of the particle, we plot it in Fig. \ref{Energy3D} for various values of $m$ and large $\Phi$.

In particular, when $\xi =\Phi =0$, the condition (\ref{cndA}) assumes the form $\gamma -\alpha -2=4\bar{n}$, where $\bar{n}$ is a positive integer. The energies in this case are given by
\begin{align}
E_{\bar{n},m}=(2\bar{n}+{\left\vert m\right\vert}&+m+1) \Omega
\notag \\
&\pm \sqrt{\Omega^{2}\left(2\bar{n}+{\left\vert m\right\vert}+m+1\right) ^{2}+M^{2}},\label{Eq_Energy_Extern_Null}
\end{align}
which is the same energy of the particle obtained in \cite{EPJC.2014.74.2935}. It should also be noted that when $M=0$ in Eq. (\ref{Eq_Energy_Extern_Null}), we obtain the Landau analog quantization. When we compare the ground state ($\bar{n} = 0$) of the result in Eq. (\ref{Eq_Energy_Extern_Null}) (see Fig. (\ref{Energy_Extern_Null})) with the energy levels of the Fig. \ref{Enery_null_const_flux}(b), which shows a symmetry between the energy of the particle and the anti-particle, we observe that such symmetry is no longer present. For negative values of $m$, the difference between energy levels is the same, whereas for positive values, the energy levels of the particle increase while that of the antiparticle is approximately constant.

\section*{Conclusion}
\label{sec:sec6}

We have presented a detailed study on the ground state of a Klein-Gordon
particle in the presence of external fields in a G\"{o}del-type spacetime. Through minimal substitution, we implemented both the Aharonov-Bohm potential and a Coulomb-type potential. The radial equation of motion was derived by means of an appropriate ansatz, and it was shown that such equation is expressed in terms of the biconfluent Heun differential equation. The expression for the ground state energy was obtained using the relations (\ref{cndA}) and (\ref{cndB}), which reveals the possibility of establishing a quantum condition between the energy of the particle and the parameter that characterizes the vorticity $\Omega $ of the spacetime. As we have mentioned, expressions for such energies are obtained without major complications only for the ground state ($n=0$). We specialized to the $n=0$ case and determined the expression for the energy and wave function of the particle. We evaluated the expressions obtained and established validity criteria for them (Eqs. (\ref{cn1}) and (\ref{cn2})). The physical implications imposed by it reveals that the energy is bounded in the finite interval $\left[E_{0m}\right]_{1,2}=\pm 1$. It was also observed an inversion of magnetic flux for certain values of quantum number $m$. Finally, it was found that the fundamental state of both the particle and the antiparticle belongs to the same spectrum, and there is no channel that allows the spontaneous creation of particles.

\section*{Acknowledgments}

The author would like to thank Cleverson Filgueiras and Fabiano M. Andrade for their critical reading of the manuscript. This work was supported by CNPq, Grants No. 427214/2016-5 (Universal), No. 303774/2016-9 (PQ), and FAPEMA, Grants No. 01852/14 (PRONEM) and 01202/16 (Universal).

\bibliographystyle{spphys}

\end{document}